\documentclass[11pt]{article}
\usepackage{amssymb}
\usepackage{graphicx}
\begin{document}

\begin{center}
{\Large\bf Dissipation of dark matter}
\end{center}

\begin{center}
Hermano Velten\footnote{email: velten@physik.uni-bielefeld.de} and Dominik J. Schwarz\footnote{email: dschwarz@physik.uni-bielefeld.de} 

Fakult\"at f\"ur Physik, Universit\"at Bielefeld, Postfach 100131, 33501 Bielefeld, Germany.
\end{center}

\begin{abstract}

Fluids often display dissipative properties. We explore dissipation in the form of bulk viscosity in the cold dark matter fluid. We constrain this 
model using current data from supernovae, baryon acoustic oscillations and the cosmic microwave background. Considering the isotropic and homogeneous background only, viscous dark matter is allowed to have a bulk viscosity $\lesssim 10^7$ Pa$\cdot$s, also consistent with the expected 
integrated Sachs-Wolfe effect (which plagues some models with bulk 
viscosity). We further investigate the small-scale formation of viscous dark matter halos, which turns out to place significantly stronger constraints on 
the dark matter viscosity. The existence of dwarf galaxies is guaranteed 
only for much smaller values of the dark matter viscosity, $\lesssim 10^{-3}$ Pa$\cdot$s.

\hspace{0.4cm}

PACS numbers: 98.80-k, 95.36+x
\end{abstract}

\section{Introduction}

Modern cosmology assumes the existence of dark energy in the form of a cosmological constant 
$\Lambda$ to dominate the cosmic energy budget today. Dark energy is believed to cause the current 
accelerated expansion of the Universe. The matter components of the Universe are baryons, photons, neutrinos and cold dark matter (CDM), which are modelled as independent and dissipationless fluids. 

This concordance model assumes from the very beginning the cosmological principle and relies on the validity of the first law of thermodynamics. This means that directional dissipative processes, i.e. heat conduction, diffusion and shear stresses, are excluded due to the high degree of isotropy imposed to the Universe. However, bulk viscosity, which is induced by a divergence of the velocity field, is the unique viscous mechanism allowed for cosmic fluids even in a homogeneous and isotropic background. Hence, the effective pressure of any cosmic fluid can be written as the sum of the kinetic pressure, $p_{\rm k}$, and the bulk viscous pressure, $p_{\rm v}$, such as ($c=1$)
\begin{equation}\label{eq1}
p_{\rm eff}= p_{\rm k} + p_{\rm v} = w_{\rm k} \epsilon - 3H\xi,
\end{equation}
where the coefficient of bulk viscosity $\xi$ is a non-negative quantity $\xi \geq 0$, due to the second law of thermodynamics \cite{weinberg}. It is obvious that the bulk viscosity reduces the effect of the kinetic pressure and can even make the effective pressure negative (see e.g.~\cite{viscous1}). Assuming (\ref{eq1}), the continuity equation of a bulk viscous fluid reads
\begin{equation}
\dot{\epsilon}+3H(\epsilon+p_k)-9H^2 \xi=0.
\end{equation}

The purpose of this paper is to investigate whether observations allow the presence of bulk viscosity in the dark matter fluid.

Clusters of galaxies revealed the existence of dark matter in the 1930s. Since then, observational evidence has shown that dark matter behaves as a nonrelativistic fluid,
$|p_{\rm{dm}}| \ll \epsilon_{\rm{dm}}$ \cite{bertone}. However, due to the lack of direct detection room is left for approaches where $p_{\rm{dm}}\neq0$ \cite{DMEoS}. The dark matter equation of 
state can be inferred using gravitational lensing and kinematic techniques at astrophysical scales as proposed by Faber and Visser \cite{faber}. Relaxing the assumption $p_{\rm dm}=0$, dark matter 
in clusters of galaxies exhibits a negative equation of state parameter $w_{\rm dm}<0$ 
\cite{mariano}. For example, the Coma galaxy cluster, one of the four clusters analyzed in 
\cite{mariano}, would be consistent with $w_{\rm dm}\sim -0.2$. Similar constraints on the constant dark matter equation of state parameter can also be found at galactic \cite{Bharadwaj} and cosmological scales \cite{EoSDMnegative}.

If $w_{\rm{dm}}$ is negative enough to allow for the observed accelerated expansion of the Universe, one could abdicate dark energy. This is the idea behind unified dark datter (UDM) models. The UDM fluid guarantees a dark matter like behavior at high redshifts, where structure formation starts, and acts as a dark energy fluid at late times. The Chaplygin gas \cite{chaplygin1}
and the viscous dark fluid (VDF) \cite{viscous1}, with their ``exotic'' equation of state, realize this idea. Both models share similar background dynamics \cite{Szy}, whereas density perturbations of the Chaplygin gas are adiabatic (as CDM) and the VDF is intrinsically nonadiabatic \cite{winfried}. For this reason the Chaplygin gas mimics more properly cold dark matter, though it is phenomenologically motivated only. On the other hand, bulk viscosity is a well-known 
dissipative phenomenon in nature. Since real fluids are always subject to dissipation, it seems reasonable to assign this property to dark matter as well. 

The UDM models with bulk viscous pressure are competitive when confronted with supernovae Ia, baryon acoustic oscillations, age constraints and expansion rate measurements \cite{backgroundbulk} and the matter power spectrum \cite{winfried}. However, the integrated Sachs-Wolfe (ISW) effect of such models severely plagues models dominated by the VDF 
\cite{barrow,oliver,dominik}. The reason of such pathology is very simple: the bulk viscosity induces a large time variation of the gravitational potential at late times. 

In this work, we propose a viscous model that remains within the 
$\Lambda$CDM conception, where $\Lambda$ drives the accelerated expansion of the Universe. We assume that CDM behaves as a real fluid equipped with bulk viscosity, viscous cold dark matter (vCDM). Within this approach we expect that the viscous effects do not cause a huge ISW effect, because (i) today's vCDM fractional energy density $\Omega_{\rm v0}\sim 0.25$ (for the UDM model $\Omega_{\rm v0}\sim 1$) and (ii) the equation of state of the viscous component assumes small negative values (in the UDM scenario we have $w_{\rm v 0}\sim -1$).

In the next section we present the dynamics of the $\Lambda$vCDM model. Its background evolution is compared with current observational data in order to find the allowed viscosity of the vCDM component. The perturbative dynamics is also investigated. We calculate the ISW effect assuming the viscosity allowed by the background analysis. We also study the growth of subhorizon vCDM perturbations leading to the formation of dark matter halos during the matter dominated epoch. In some sense, we extend the findings concerning viscous UDM of our previous work \cite{dominik}. We discuss our results in the last section. 

\section{Dynamics of the $\Lambda$vCDM model and observational constraints}

\subsection{Background Dynamics}

The isotropic and homogeneous background dynamics of the $\Lambda$vCDM model resembles that of the flat $\Lambda$CDM model. The Hubble expansion rate $H$ 
is given in terms of the fractional energy densities $\Omega_i$, where the subscript $i$ stands for baryons (b), radiaton (r), vCDM (v) and cosmological constant ($\Lambda$),
\begin{equation}
H^{2}=H^{2}_{0}\left[\Omega_{\rm b0}(1+z)^3+\Omega_{\rm r0}(1+z)^4+\Omega_{\rm v}(z)+\Omega_{\Lambda}\right].
\label{H}
\end{equation}
In order to obtain the function $\Omega_{\rm v}\left(z\right)$ we have to specify the pressure of the vCDM and solve its continuity equation. 

Within relativistic thermodynamics the comoving frame is related to either the energy transport (Landau frame \cite{landau}) or to the particle number transport (Eckart frame \cite{eckart}). Here, we adopt the latter approach. For the homogeneous and isotropic background, dissipative effects manifest
itself only through bulk viscosity. Shear viscosity and heat conduction could be relevant effects at the perturbative level, but we do not include them in this study. We assume that the vCDM fluid has a vanishing kinetic pressure and an intrinsic bulk viscous pressure
\begin{equation}
p_{\rm v}=-3 H \xi.
\end{equation} 
The choice of $\xi$ seems to be the crucial aspect of any viscous model. In this work, we follow the recent results obtained in \cite{barrow,oliver,dominik} and stick to the same choice for the bulk viscous coefficient
\begin{equation} 
\xi=\xi_0\left(\frac{\epsilon_{\rm v}}{\epsilon_{\rm  v0}}\right)^{\nu}, 
\label{xi}
\end{equation} 
where $\xi_0$ and $\nu$ are constants and $\epsilon_{\rm v0}$ is the density of the vCDM fluid today. This means that the current viscosity of such dark fluid is given by the parameter $\xi_0$. 

For viscous UDM models, the ansatz (\ref{xi}) can lead to a large amplification of the ISW signal \cite{barrow,oliver,dominik}. Fixing 
$\nu=0$ or $\nu=-1/2$, the ISW effect problem of these viscous cosmologies is less severe \cite{dominik}. The quoted values for $\nu$ have an explicit physical interpretation: the former means a constant bulk viscosity and the latter implies, for the one-fluid approximation (i.e.\ the bulk viscous fluid corresponds to the total density), the same background dynamics as seen in the $\Lambda$CDM model. We study the cases $\nu=0$ and $\nu=-1/2$ and refer to them as models A and B, respectively.

For model A, the continuity equation for the vCDM fluid reads
\begin{equation}
(1+z)\frac{d \Omega_{\rm v}(z)}{dz}-3\Omega_{\rm v}\left(z\right)+\tilde{\xi}\left[\Omega_{\rm r0}(1+z)^{4}+\Omega_{\rm b0}(1+z)^{3}+\Omega_{\rm v}\left(z\right)+\Omega_{\Lambda}\right]^{1/2}=0,
\label{model1}
\end{equation}
and for model B it becomes
\begin{equation}
(1+z)\frac{d \Omega_{\rm v}(z)}{dz}-3\Omega_{\rm v}\left(z\right)+\tilde{\xi}\Omega_{\rm v0}^{1/2}\Omega_{\rm v}\left(z\right)^{-1/2}\left[\Omega_{\rm r0}(1+z)^{4}+\Omega_{\rm b0}(1+z)^{3}+\Omega_{\rm v}\left(z\right)+\Omega_{\Lambda}\right]^{1/2}=0,
\label{model2}
\end{equation}
where the definition
\begin{equation} \label{tildexi}
\tilde{\xi}=\frac{24\pi G \xi_0}{H_0}
\end{equation}
is valid for both models. The vCDM equation of state parameter $w_{\rm v}=p_{\rm v}/\epsilon{\rm v}$ today is

\begin{equation} \label{wv0}
w_{\rm v 0}=-\frac{\tilde{\xi}}{3 \Omega_{\rm v 0}}.
\end{equation}

Note that the CDM model is recovered if $\tilde{\xi}=0$. Fixing the values $\Omega_{\rm b0}=0.043$ and $\Omega_{\rm r0}=8.32\times 10^{-5}$ as given by the WMAP seven-year data \cite{komatsu11}, the remaing free parameters of our 
viscous models are $\tilde{\xi}$ and $\Omega_{\Lambda}$ (with $\Omega_{\rm v0}=1-\Omega_{\rm b0}-\Omega_{\rm r0}-\Omega_{\Lambda}$).

\subsection{Metric and density perturbations}

We assume a homogeneous, isotropic and spatially flat Universe, where the line element for scalar perturbations in the Newtonian gauge without anisotropic stress is
\begin{eqnarray}
ds^{2}=a^{2}\left(\eta\right)\left[-\left(1+2\psi\right)d\eta^{2}+\left(1-2\psi\right)\delta_{ij}dx^{i}dx^{j}\right].
\end{eqnarray}

For convenience we introduce $\mathcal{H}=a^{\prime}/a$, where prime means derivative w.r.t.\ the conformal time $\eta$. Let us first calculate the ISW effect [a net change in the energy of cosmic microwave background (CMB) photons produced by evolving potentials wells] of our viscous models. Given the gravitational potential $\psi$ it can be calculated by
\begin{equation}
\left(\frac{\Delta T}{T}\right)_{ISW}=2 \int^{\eta_{0}}_{\eta_{r}} \psi^{\prime} d\eta ,
\label{ISW}
\end{equation}
where the integration is performed along the photon trajectory from the conformal time at the recombination ($\eta_{r}$) to the conformal time today ($\eta_{0}$).

The momentum constraint reads
\begin{eqnarray}
-k^{2}\psi-3\mathcal{H}\psi^{\prime}-3\mathcal{H}^{2}\psi = \frac{3H^{2}_{0}a^2}{2}\left\{\Omega_{\rm b}\Delta_{\rm b}+\Omega_{\rm v}\Delta_{\rm v}\right\},
\label{poisson}
\end{eqnarray}
where, for each component, we have defined the energy density contrast $\Delta=\delta \epsilon / \epsilon$. Since we are interested in the late time ISW effect, we have neglected the perturbations in the radiation fluid ($\Omega_r\Delta_r = 0$).

Additionally to (\ref{poisson}) we have the remaining perturbed Einstein equations
\begin{eqnarray}
-k\left(\psi^{\prime}+\mathcal{H}\psi\right)=\frac{3H^{2}_{0}a^2}{2}\left\{\Omega_{\rm b}\theta_{\rm b}+(1+w_{\rm v})\Omega_{\rm v}\theta_{\rm v}\right\},
\label{Einstein2}
\end{eqnarray}
\begin{equation}
\psi^{\prime\prime}+3\mathcal{H}\psi^{\prime}+(2\mathcal{H^{\prime}}+\mathcal{H}^{2})\psi=\frac{3a^2H^{2}_{0}\Omega_{v}}{2}
\left[-\frac{w_{\rm v}}{3\mathcal{H}}\left(k \theta_{\rm v} +3\mathcal{H}\psi+3\psi^{\prime}\right)+\nu w_{\rm v}\Delta_{\rm v}\right],
\label{Einstein3}
\end{equation}
where for each fluid we have defined the scalar velocity perturbation $\theta$ by means of $\delta u^{j}_{;j}=-k \theta/a$ with $k$ being the wave number. 

Since there is no interaction between the different components of our model, each fluid obeys separately the perturbed energy-momentum conservation equations, $\delta T^{\mu}_{\alpha;\mu}=0$ (see \cite{dominik} for a detailed derivation), which are used to close the set of equations. 

This set of equations must be solved numerically. Once we obtain the function $\psi$, the ISW effect can be computed using (\ref{ISW}).  

Cosmological perturbation theory also allows us to investigate the formation of dark matter halos that attract baryons in order to form galaxies. The standard cosmological scenario gives rise to a hierarchical formation process where small structures form first. vCDM behaves as a pressureless fluid at the beginning of the matter dominated phase, i.e., $w_{\rm v} (z \gg 0)\sim 0$. However, the equation of state evolves in time departing from the pressureless behavior as the Universe expands. Additionally, there are also nonadiabatic contributions to the dynamics of density fluctuations. 
Assuming (\ref{xi}), the subhorizon vCDM perturbations obey the following Meszaros-like equation (see \cite{dominik} for details):
\begin{equation}
a^{2}\frac{\,d^{2}\Delta_{\rm v}}{da^{2}}+\left[\frac{a}{H}\frac{d
\,H}{da}+3+A(a)+B(a)k^{2}\right]a\frac{\,d\Delta_{\rm v}}{da}+\left[+C(a)+D(a)k^{2}-\frac{3}{2}\right]\Delta_{\rm v}=P(a),
\label{small}
\end{equation}
\begin{eqnarray}
A(a)=-6w_{\rm v}+\frac{a}{1+w_{\rm v}}\frac{dw_{\rm v}}{da}-\frac{2a}{1+2w_{\rm
v}}\frac{dw_{\rm v}}{da}+\frac{3w_{\rm v}}{2(1+w_{\rm v})}\nonumber
\end{eqnarray}
\begin{eqnarray}
B(a)=-\frac{w_{\rm v}}{3a^{2}H^{2}(1+w_{\rm v})}\nonumber
\end{eqnarray}
\begin{eqnarray}
C(a)=\frac{3w_{\rm v}}{2(1+w_{\rm v})}-3w_{\rm v}-9w^{2}_{\rm v}-\frac{3w^{2}_{\rm
v}}{1+w_{\rm v}}\left(1+\frac{a}{H}\frac{dH}{da}\right)-3a\left(\frac{1+2w_{\rm
v}}{1+w_{\rm v}}\right)\frac{dw_{\rm v}}{da}+\frac{6aw_{\rm v}}{1+2w_{\rm
v}}\frac{dw_{\rm v}}{da}\nonumber
\end{eqnarray}
\begin{eqnarray}
D(a)=\frac{w^{2}_{\rm v}}{a^{2}H^{2}(1+w_{\rm v})}\nonumber
\end{eqnarray}
\begin{eqnarray}
P(a)= - 3 \nu w_{\rm v} a \frac{d\Delta_{\rm v}}{da} + 
3\nu w_{\rm v} \Delta_{\rm v} \left[-\frac{1}{2} + \frac{9w_{\rm v}}{2} +\frac{-1-4w_{\rm
v}+2w_{\rm v}^2}{w_{\rm v}(1+ w_{\rm v})(1+2w_{\rm v})} a \frac{d\,w_{\rm v}}{da} - 
\frac{ k^{2}(1-w_{\rm v})}{3H^{2}a^{2}(1+w_{\rm v})}\right]\nonumber
\end{eqnarray}
where the scale factor was used as the dynamical parameter. The function $P(a)$ contains the contributions from the perturbation of $\xi$. Thus, for model A ($\nu=0$) we have $P(a)=0$. A remarkable point here is that the evolution of subhorizon vCDM perturbations is scale dependent. As shown in \cite{dominik}, a viscosity large enough to accelerate the current Universe leads to a substantial suppression of growth at small scales.

\subsection{Comparison with observations}

In Fig.~1 we show constraints on the free parameters $\tilde{\xi}$ and $\Omega_{\Lambda}$ from different observational data sets. We constrain the background dynamics of our viscous models by means of the following observations: supernovae (SN) data (here we have used the Constitution sample \cite{SN}), the baryon acoustic oscillation (BAO) parameter $A(z)$ \cite{BAO} from the WiggleZ Dark Energy Survey \cite{wiggleZ}, and the position of the observed CMB peak $l_1$ obtained by the WMAP project \cite{CMB} that is related to the angular scale $l_A$ \cite{wu}. The solid lines in Fig.~1 are the $2\sigma$ confidence level contours obtained from the likelihood function $\mathcal{L}\propto {\rm exp}(-\chi^{2}/2)$. The standard $\chi^{2}$ statistics, 
\begin{equation}
\chi^2 (\tilde{\xi},\Omega_{\Lambda},H_0)=\sum^{N}_{i=1}\frac{(D^{\rm th}_{i}-
D^{\rm obs}_{i})^2}{\sigma^2_i},
\end{equation}
measures the goodness of the fit. For each data set, with $N$ data points, the theoretical value obtained within the $\Lambda$vCDM model $D^{\rm th}$ is confronted with the observation $D^{\rm obs}$. In order to obtain the bidimensional likelihood contours shown in Fig.~1, we marginalize $H_0$ with a flat prior [$0<H_0 ($Km$^{-1}\, $s$\, $Mpc$) <100$].
Long dashed lines are age constraints corresponding to $13$~Gyr and 
$14$~Gyr. The gray filled areas in both panels of Fig.~1 correspond to the ``concordance'' parameter values allowed at $2\sigma$ confidence level.
The best fit occurs at $\tilde{\xi}=0$, but it is possible to establish an upper bound (at $2\sigma$) to the viscosity parameter $\tilde{\xi}$. For the model A (B) this value is $\tilde{\xi}\lesssim 0.24$ $(0.31)$ as seen in the horizontal dashed line in the left (right) panel in Fig.~1.

\begin{figure}[!t]
\begin{center}
\includegraphics[width=0.47\textwidth]{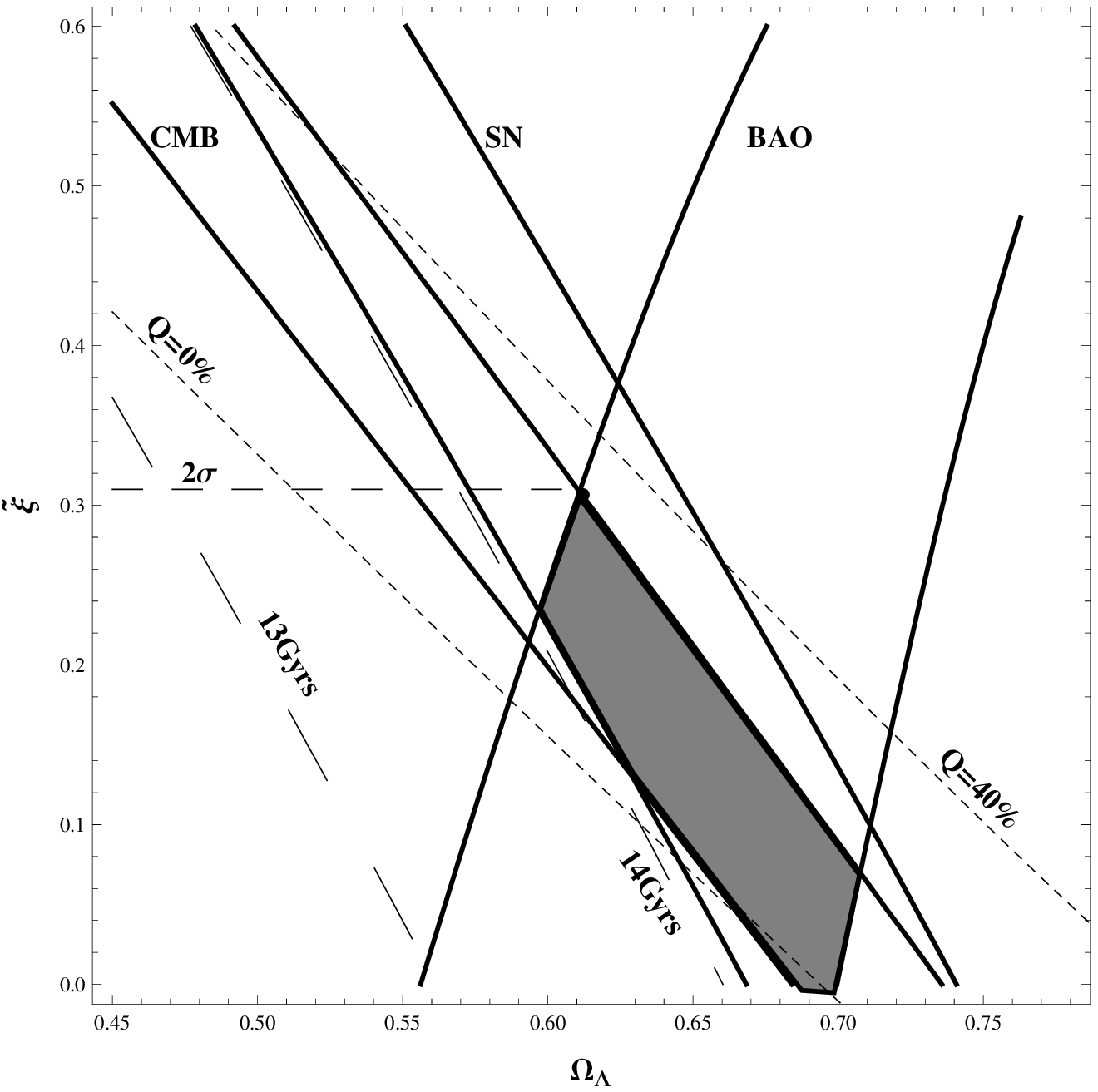}
\includegraphics[width=0.47\textwidth]{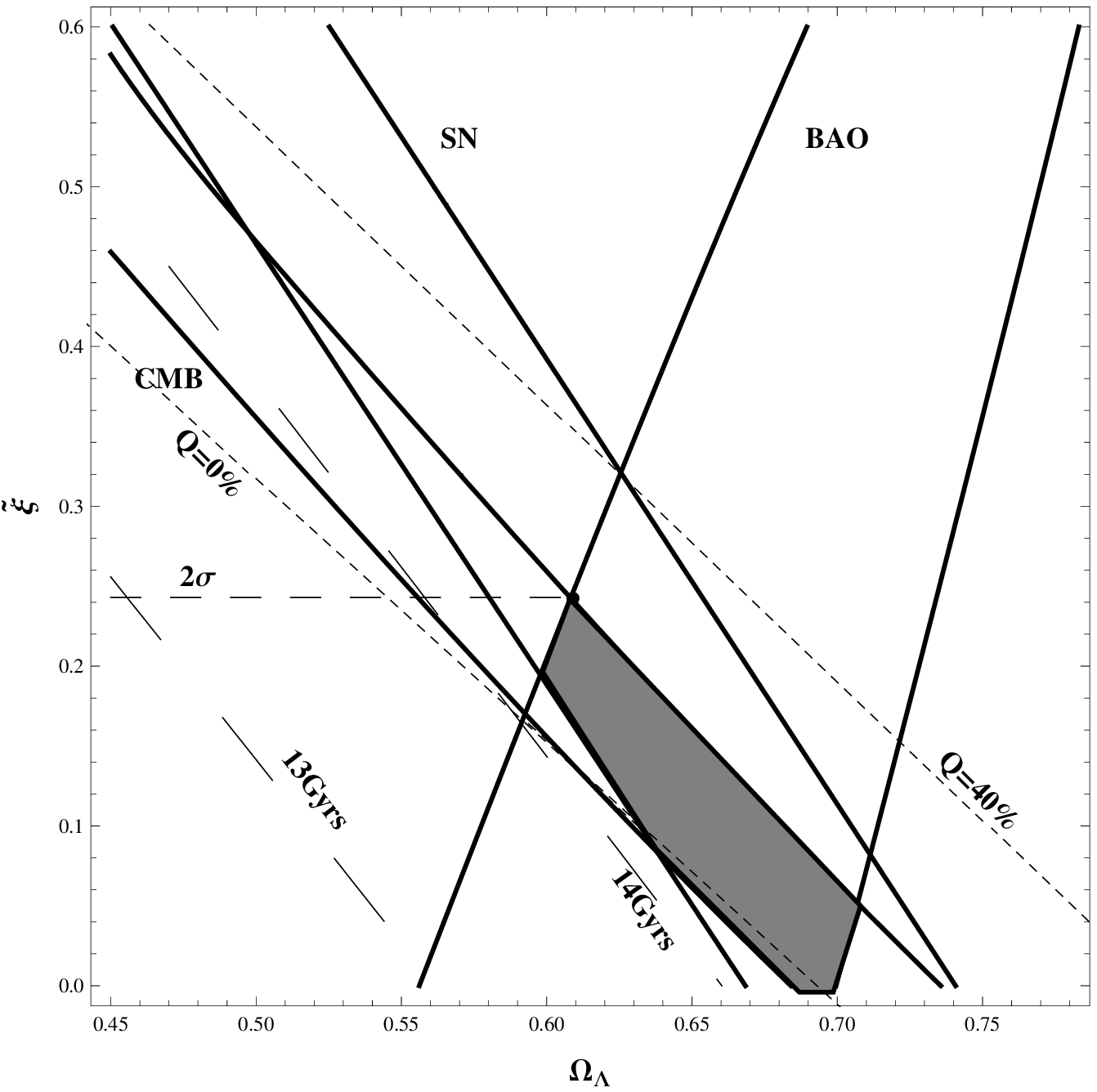}
\label{f1}	
\caption{Observational constraints on the parameter space ($\tilde{\xi}$, $\Omega_{\Lambda}$) for model A (left) and model B (right). Solid lines are the contours of $2\sigma$ confidence level. Long dashed lines are the age constraints ($13$ Gyr and $14$ Gyr). Short dashed lines, from bottom to top, correspond to $Q=0$ and $Q=40\%$ (see text for explanation). The horizontal dashed line sets the maximum allowed viscosity at $2 \sigma$.}
\end{center}
\end{figure}

With the equations for the potential, derived in Sec. 2, we compute the ISW effect of both viscous models and compare it with the prediction of a flat $\Lambda$CDM model (the fiducial cosmology adopted here has parameters 
$H_0=72$ km/s/Mpc and $\Omega_{m0}=0.266$, as suggested by the WMAP seven-year analysis \cite{komatsu11}). The short dashed lines in Fig.~1 correspond to relative amplifications ($Q$) of the ISW effect calculated as
\begin{equation}
Q\equiv\frac{\left(\frac{\Delta T}{T}\right)_{\rm ISW}^{\rm \Lambda vCDM}}{\left(\frac{\Delta T}{T}\right)_{\rm ISW}^{\rm \Lambda CDM}}-1.
\label{Q}
\end{equation}
If $Q>0 \,(<0)$ the $\Lambda$vCDM model produces more (less) temperature variation to the CMB photons via the ISW effect than the fiducial $\Lambda$CDM model. Such an analysis of the ISW effect has been proposed in \cite{dent} and used in \cite{dominik}. As seen in Fig.~1, the parameters allowed by the background correspond to values between $Q=0\%$ and $Q=40\%$. This is a reduced ISW effect, when compared with the viscous UDM models where the amplification could reach $Q \sim 120\%$ \cite{dominik}.

\begin{figure}
\begin{center}
\includegraphics[width=0.46\textwidth]{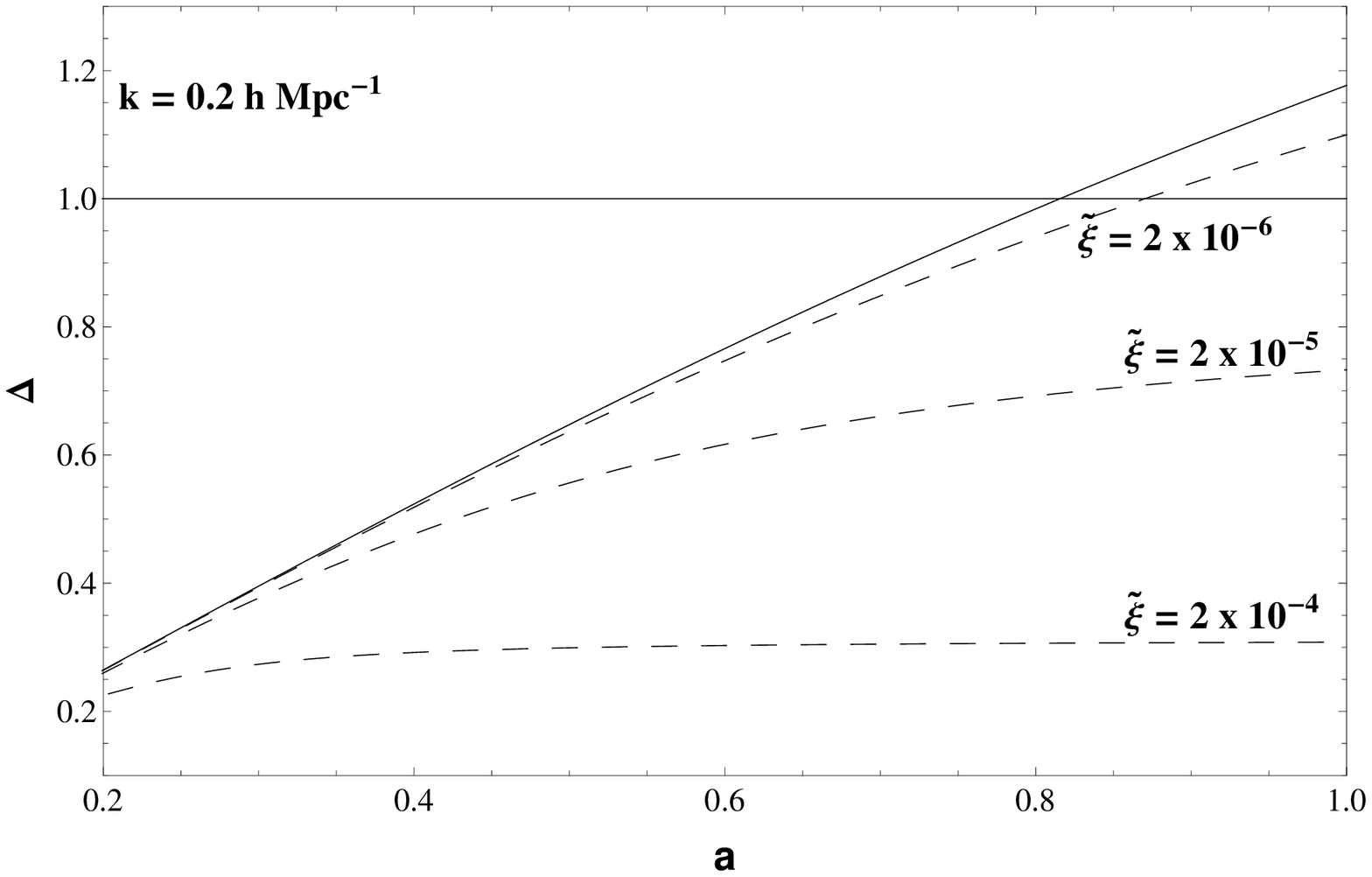}
\includegraphics[width=0.46\textwidth]{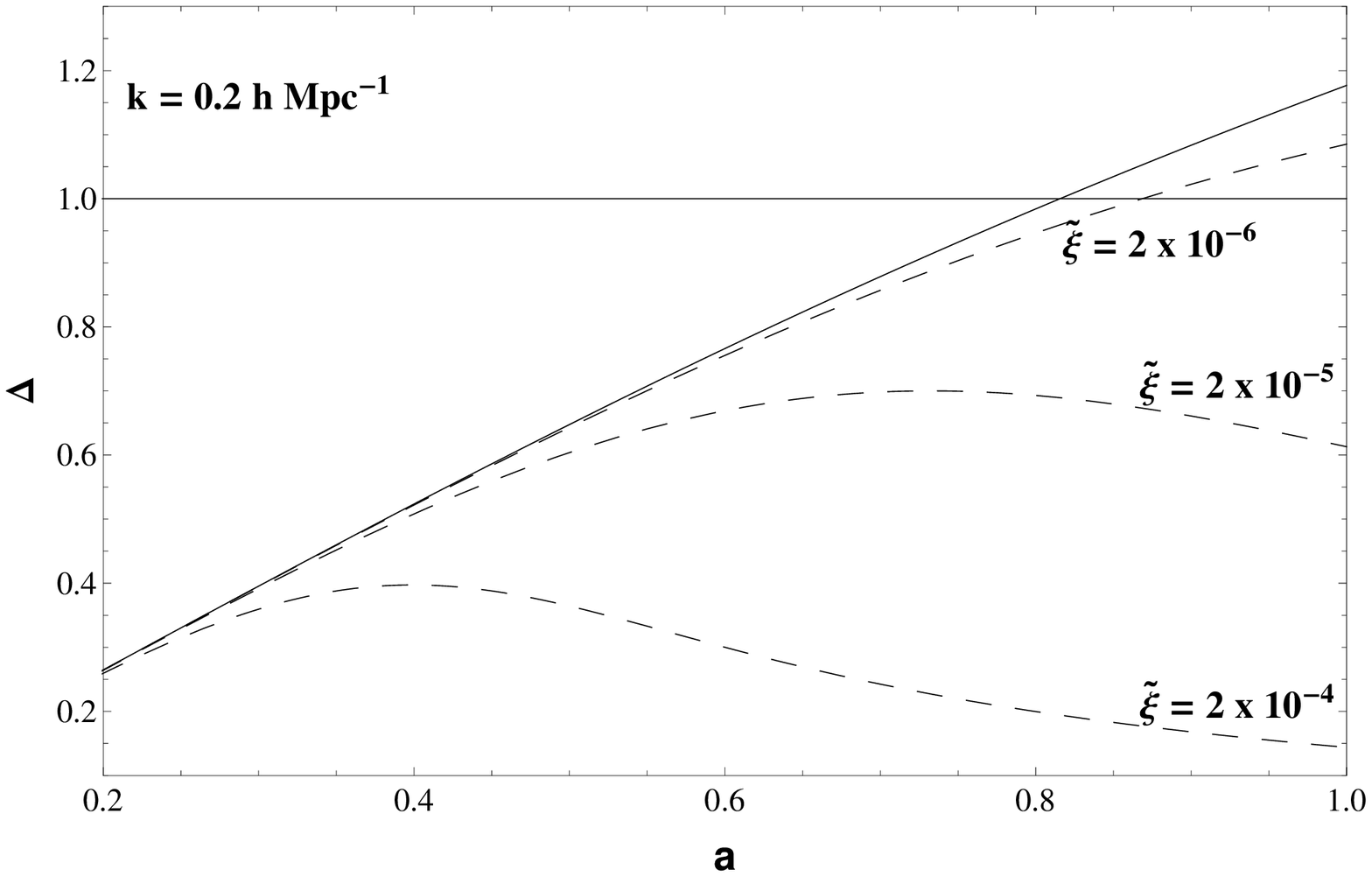}
\includegraphics[width=0.46\textwidth]{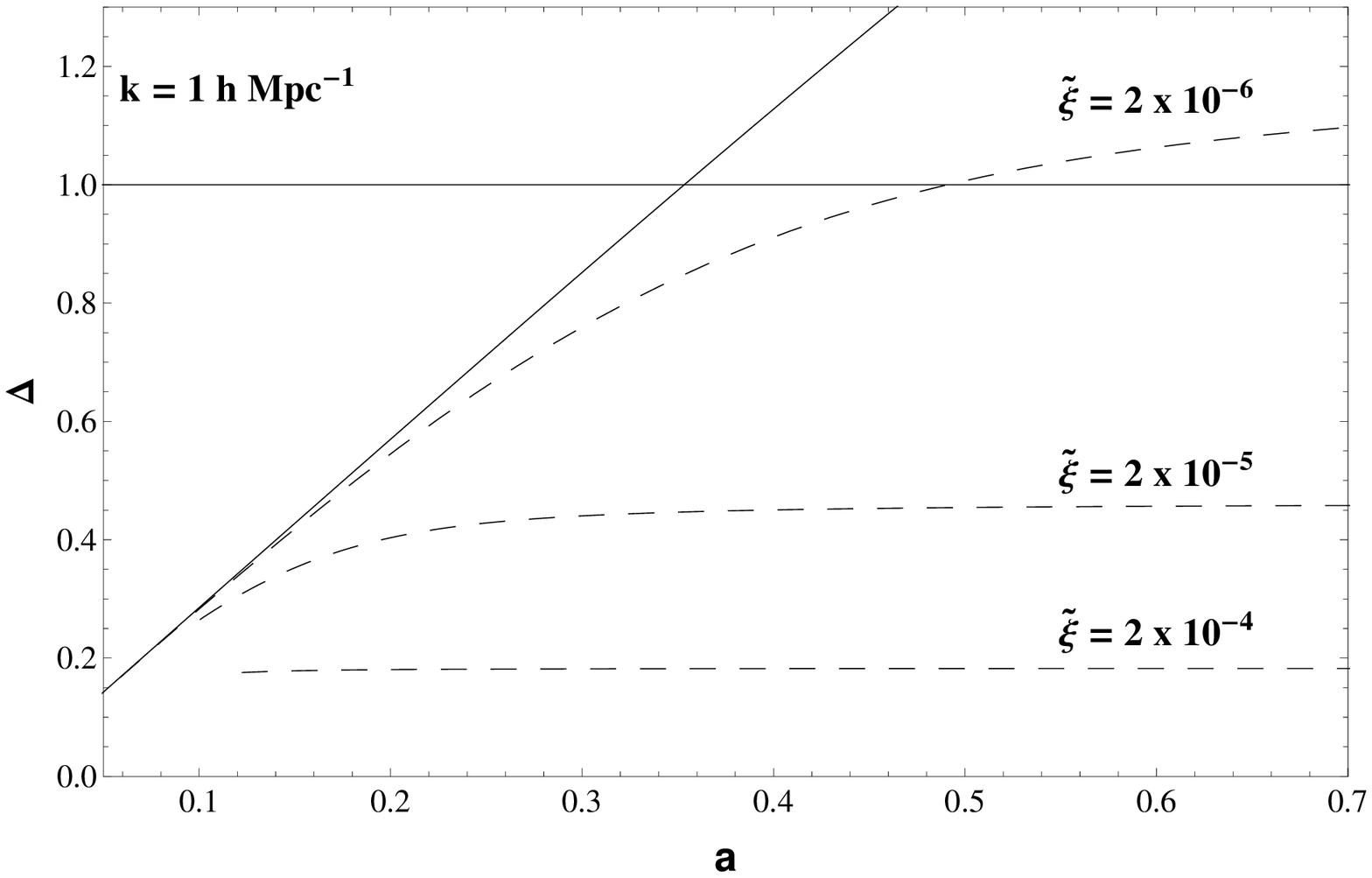}
\includegraphics[width=0.46\textwidth]{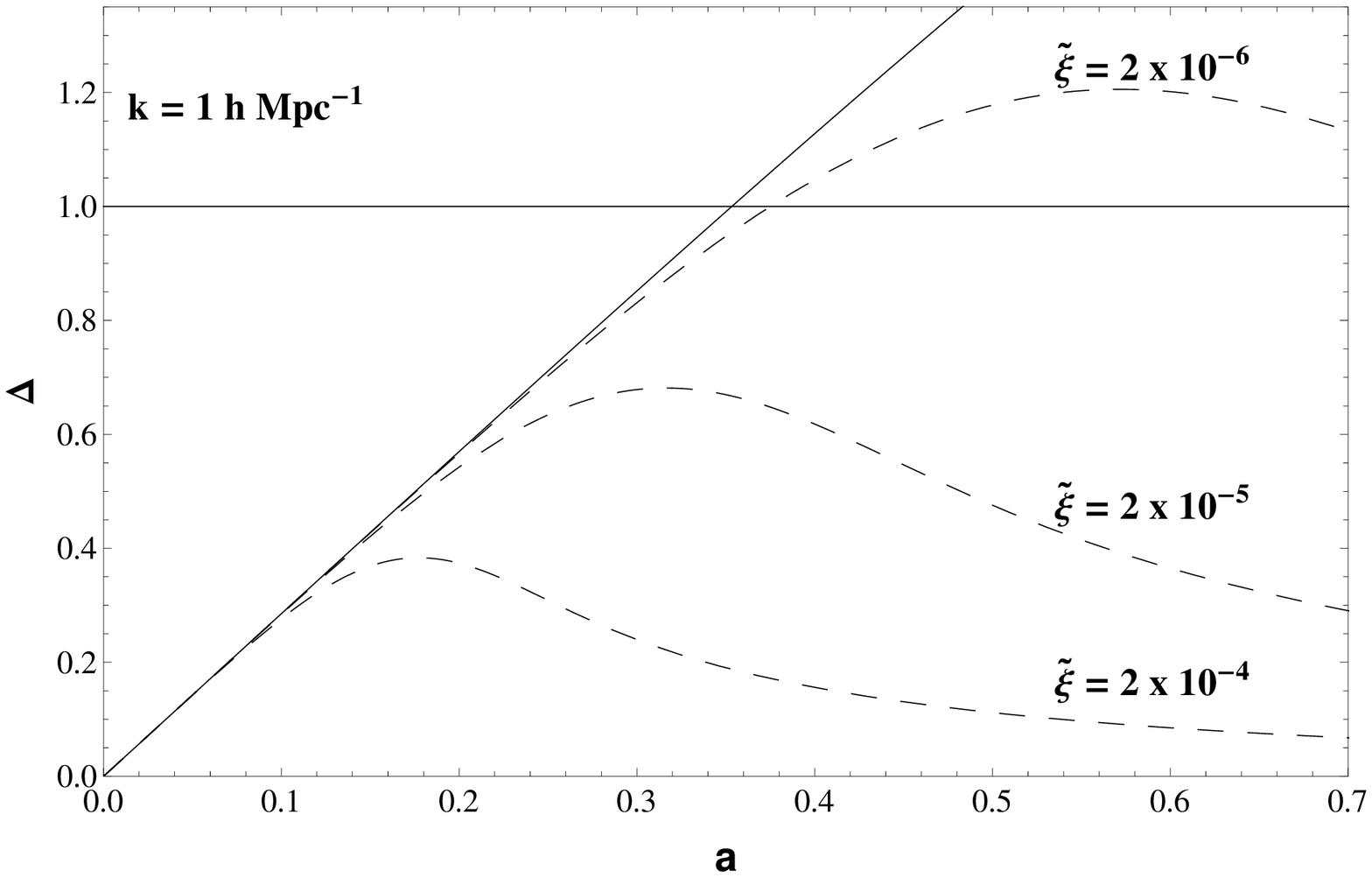}
\includegraphics[width=0.46\textwidth]{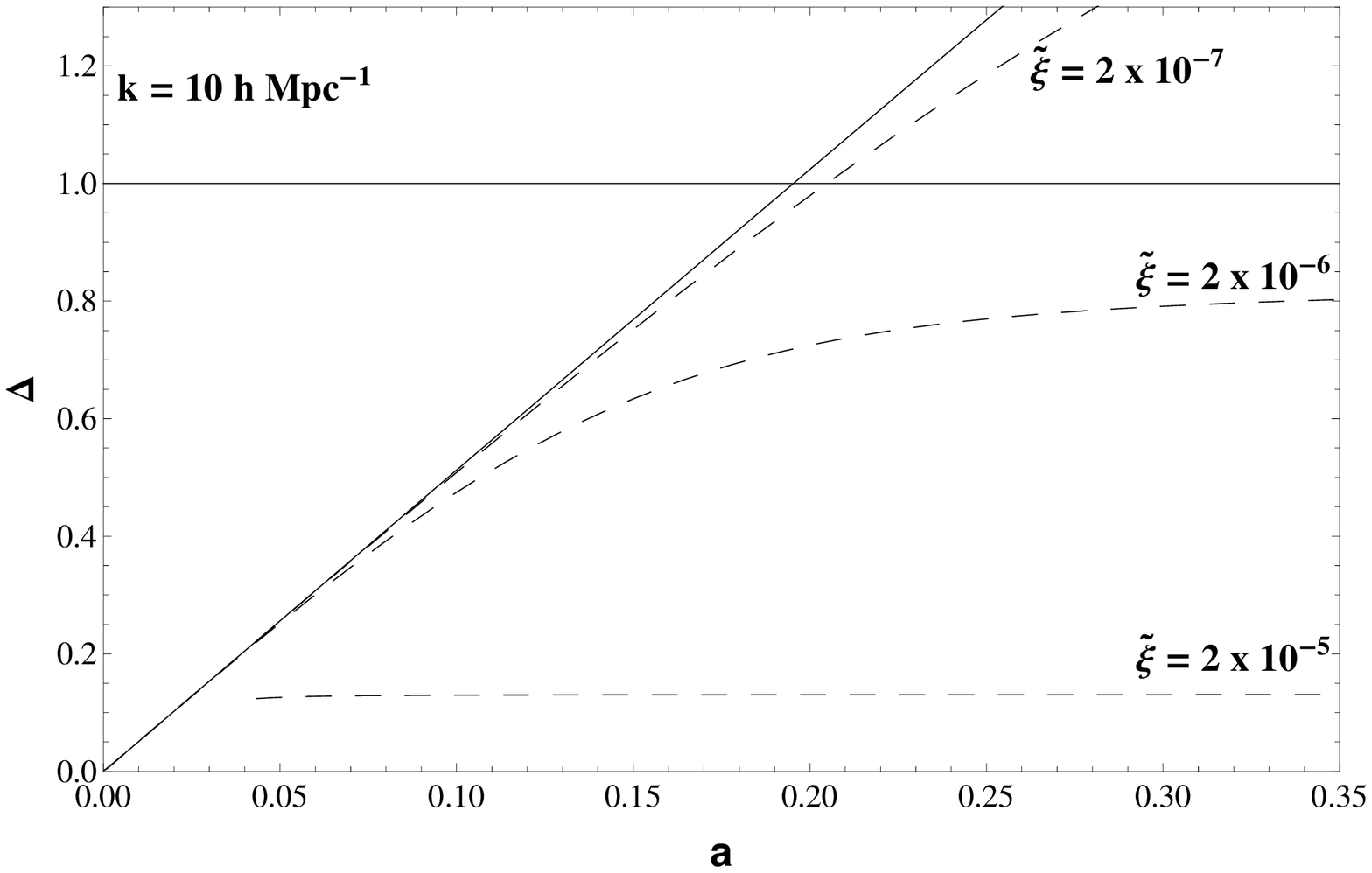}
\includegraphics[width=0.46\textwidth]{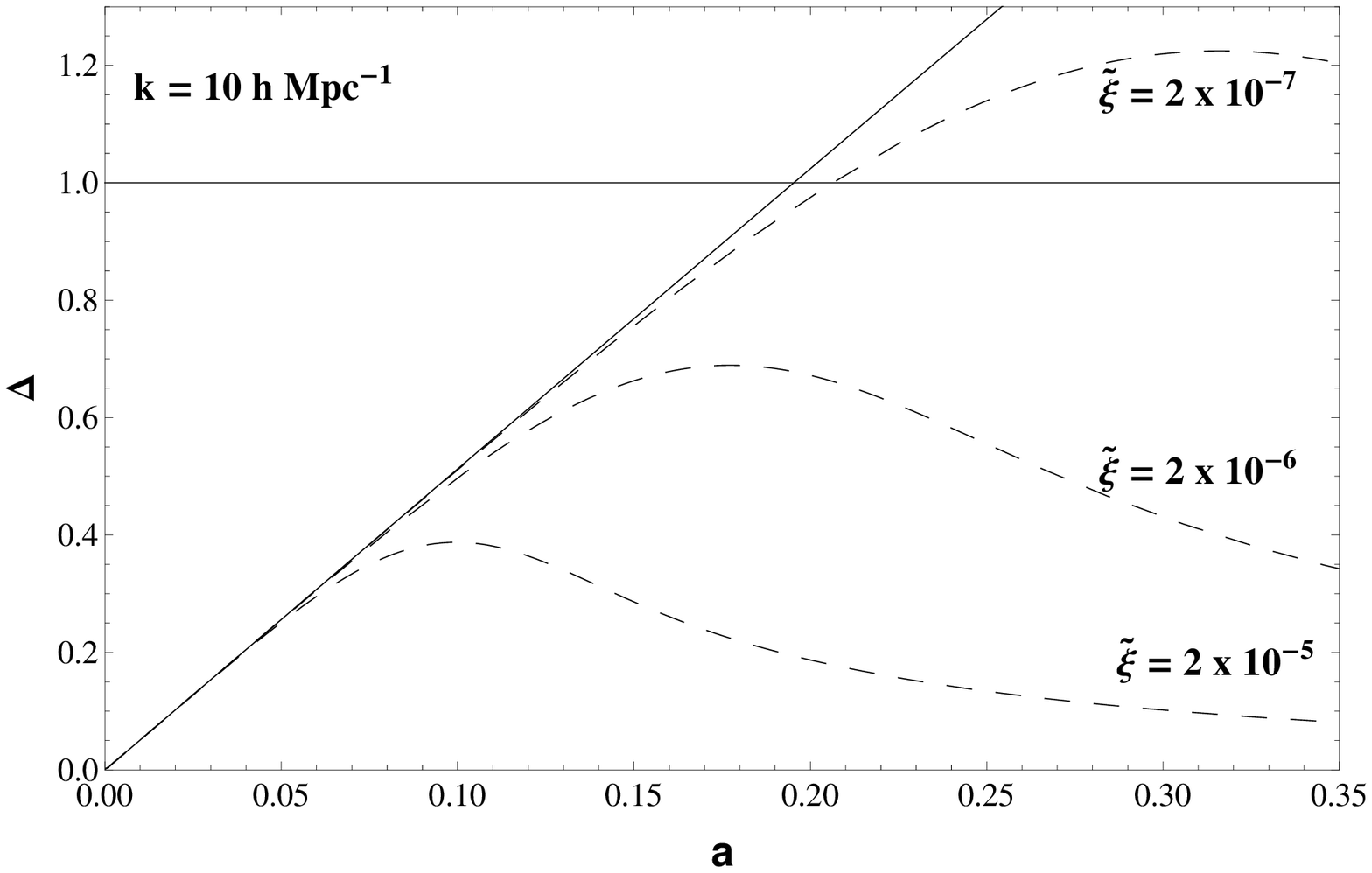}
\includegraphics[width=0.46\textwidth]{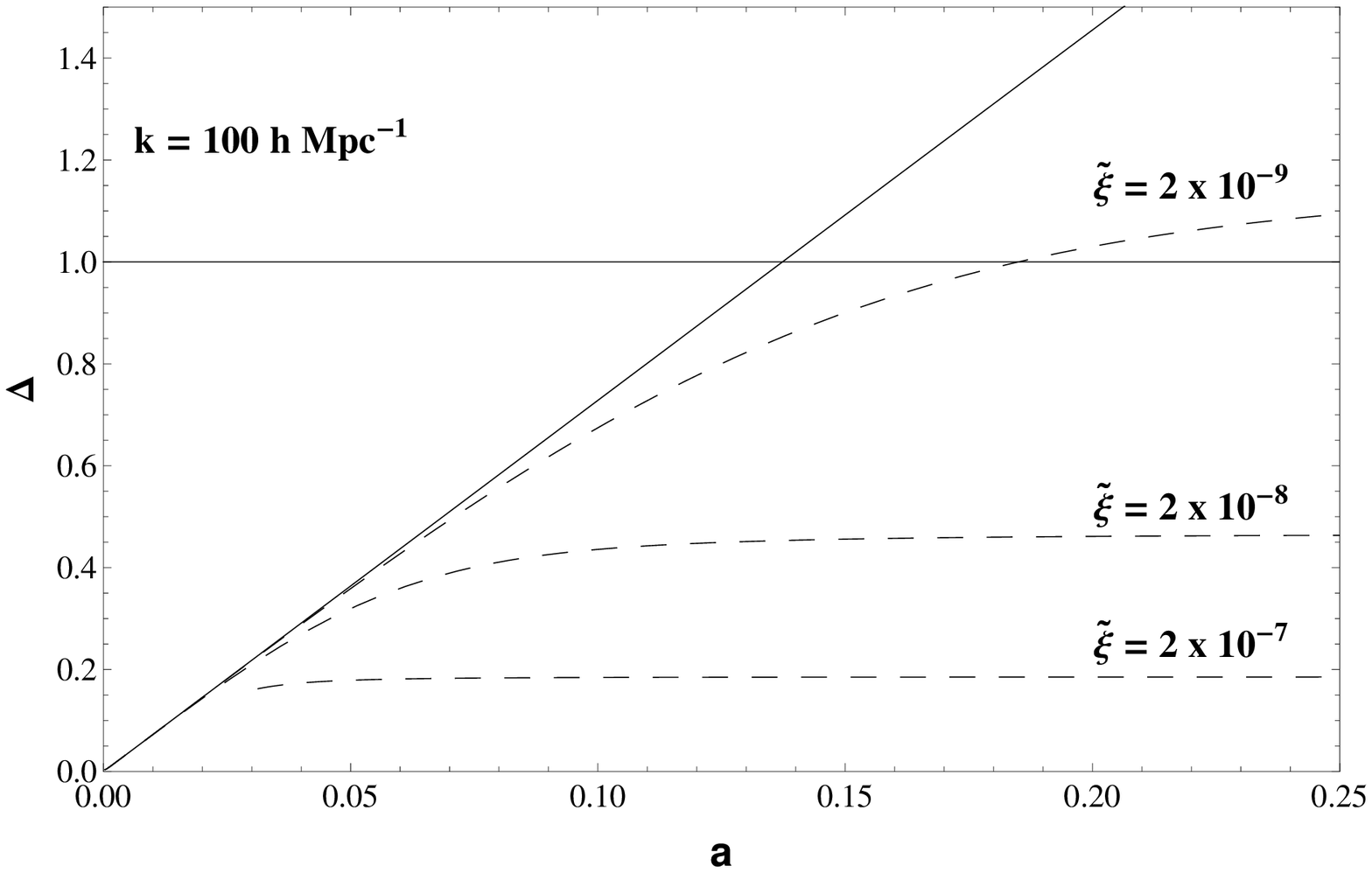}
\includegraphics[width=0.46\textwidth]{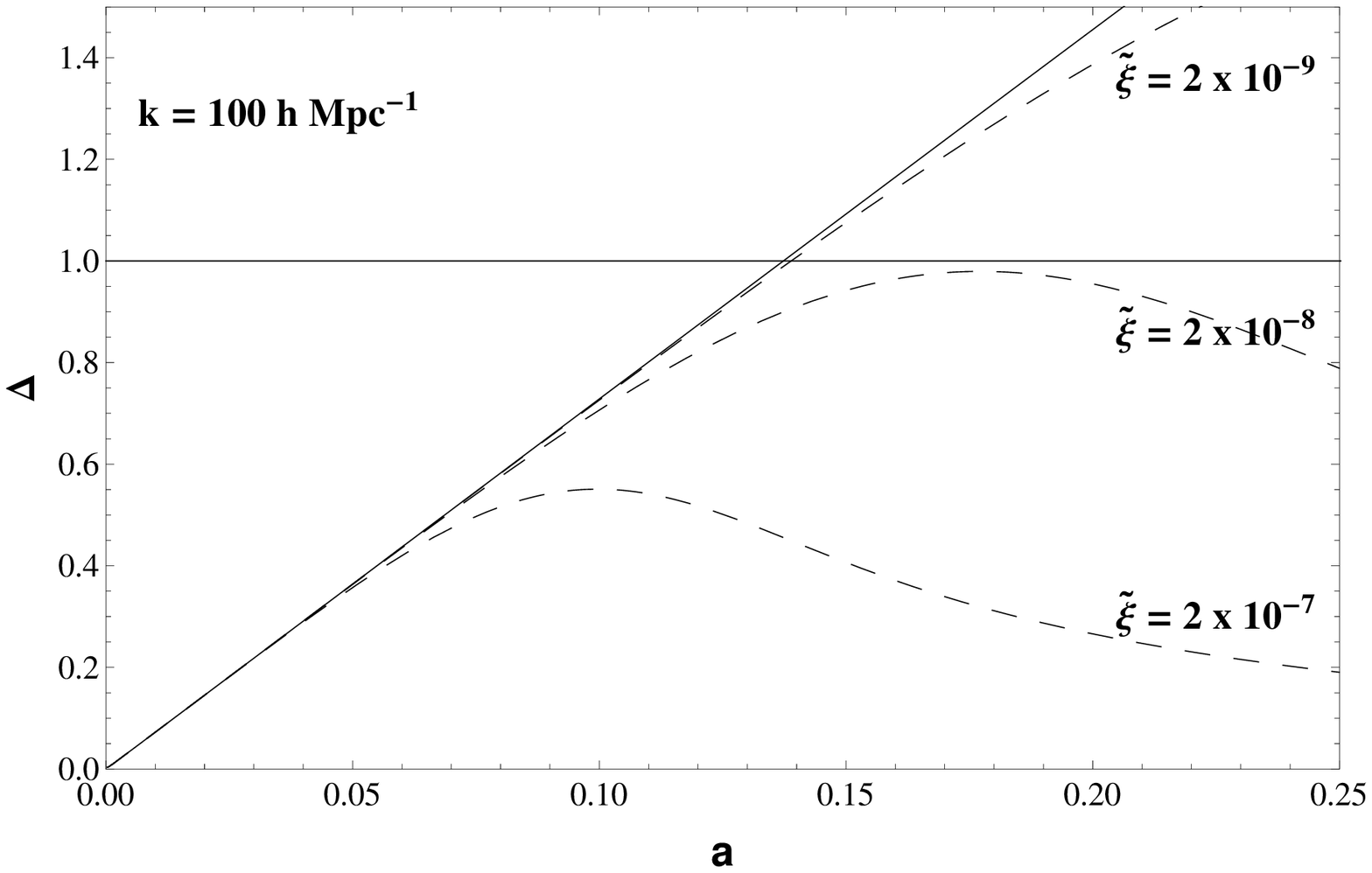}
\includegraphics[width=0.46\textwidth]{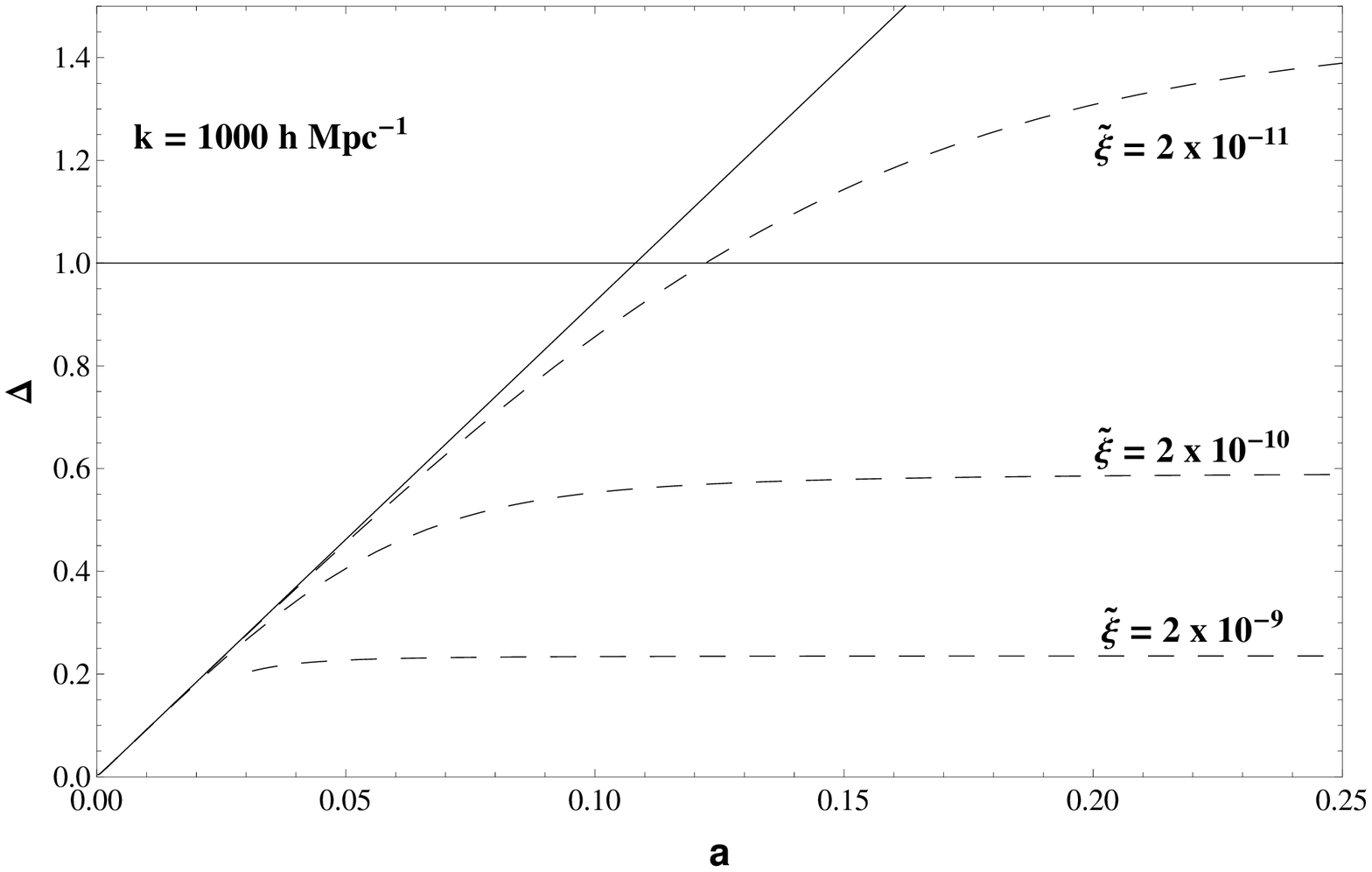}
\includegraphics[width=0.46\textwidth]{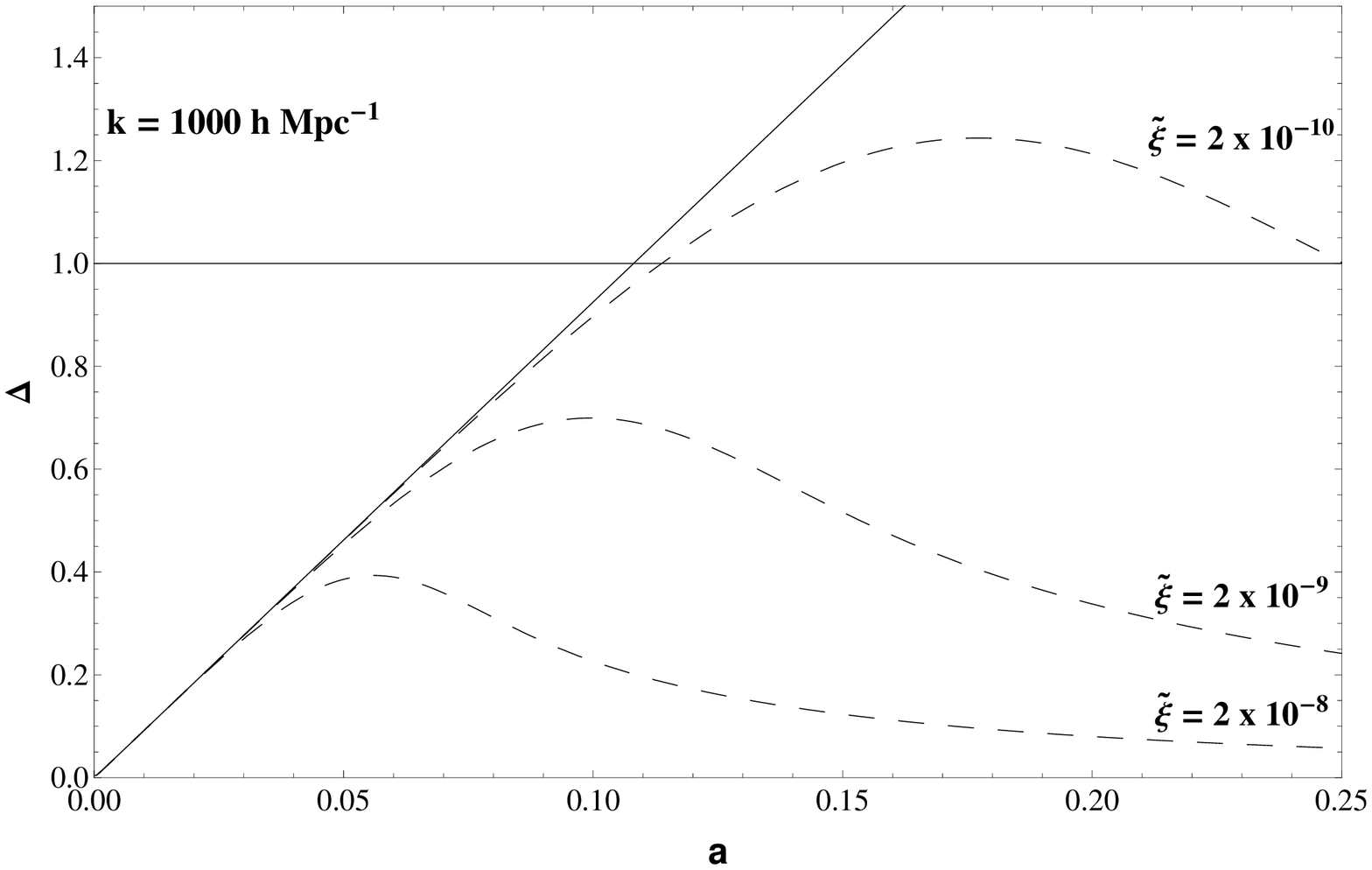}
\label{f3b}	
\caption{Growth of subhorizon density perturbations for different scales in the range $k=1000h$ ${\rm Mpc}^{-1}$ (dwarf galaxy), in the bottom panels, to $k=0.2h$ ${\rm Mpc}^{-1}$ (galaxy clusters) in the top panels. Left (right) panels correspond to model A (B). The solid line corresponds to the standard CDM $\Delta \propto a$ result. Dashed lines correspond to the viscous CDM growth for various values of the parameters $\tilde{\xi}$ as indicated in the panels. }
\end{center}
\end{figure}

A successful structure formation process (within the hierarchical scenario) is achieved when small structures (formed from the smallest halos) merge into large matter agglomerations. At the moment of matter-radiation equality 
($z_{\rm eq}\sim3000$) typical CDM subhorizon perturbations grow 
$\propto a$. Equation (\ref{small}) also has this solution in the limit $w_{\rm v}=0$. However, the vCDM ($w_{\rm v}\neq0$) perturbation evolution is scale dependent. 

In Fig.~2, we plot the evolution of the density contrast $\Delta$ for both viscous models for scales in the range $k=1000h$ ${\rm Mpc}^{-1}$ (dwarf galaxies) to $k=0.2h$ ${\rm Mpc}^{-1}$ (galaxy clusters). In each panel the solid line is the standard CDM growth $\propto a$. The dashed lines correspond to the viscous models for different values of the viscosity coefficient. The initial conditions, i.e. the power spectrum at the matter-radiation equality, are set using the CAMB code \cite{CAMB}. This provides the correct amplitude for each $k$ mode at $z_{\rm eq}$ and helps us to identify the onset of the nonlinear structure formation $\Delta (z_{\rm nl})=1$. Considering $\tilde{\xi} \sim 0.2$, as obtained above, the viscous effects would suppress the halo growth well before $z_{\rm eq}$. Viscous dark halos at cluster scales ($k= 0.2 h {\rm Mpc}^{-1}$) are able to follow the typical CDM growth only if $\tilde{\xi}\lesssim 10^{-6}$. Smaller scales place stronger constraints on the bulk viscosity, for dwarf galaxies $\tilde{\xi}\lesssim 10^{-11}$ in model A and $\tilde{\xi}\lesssim 10^{-10}$ in model B.

In Fig.~3 we implement a more conservative analysis. It shows the maximum viscosity allowed in order to reproduce the standard CDM perturbation growth until $\Delta =1$. The left panel displays the constraints on the dimensionless quantity $\tilde{\xi}$. The right panel shows the same results in SI units. The interpretation of these plots is as follows: For the wave numbers shown on the horizontal axis, linear perturbations of the vCDM fluid are identical to standard CDM if the viscosity is below the indicated critical values. For larger values, structure formation is affected.

\begin{figure}
\begin{center}
\includegraphics[width=0.49\textwidth]{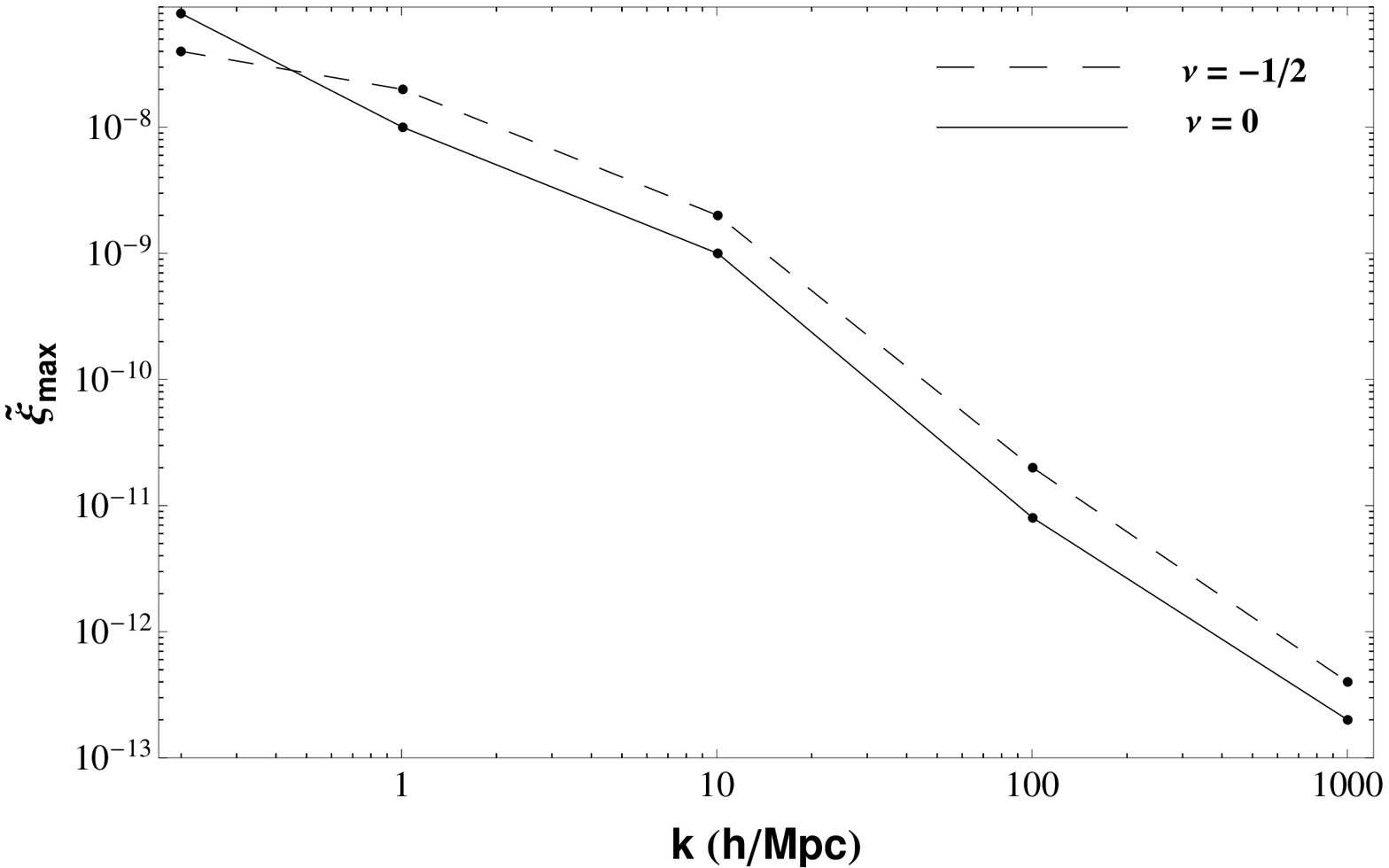}
\includegraphics[width=0.49\textwidth]{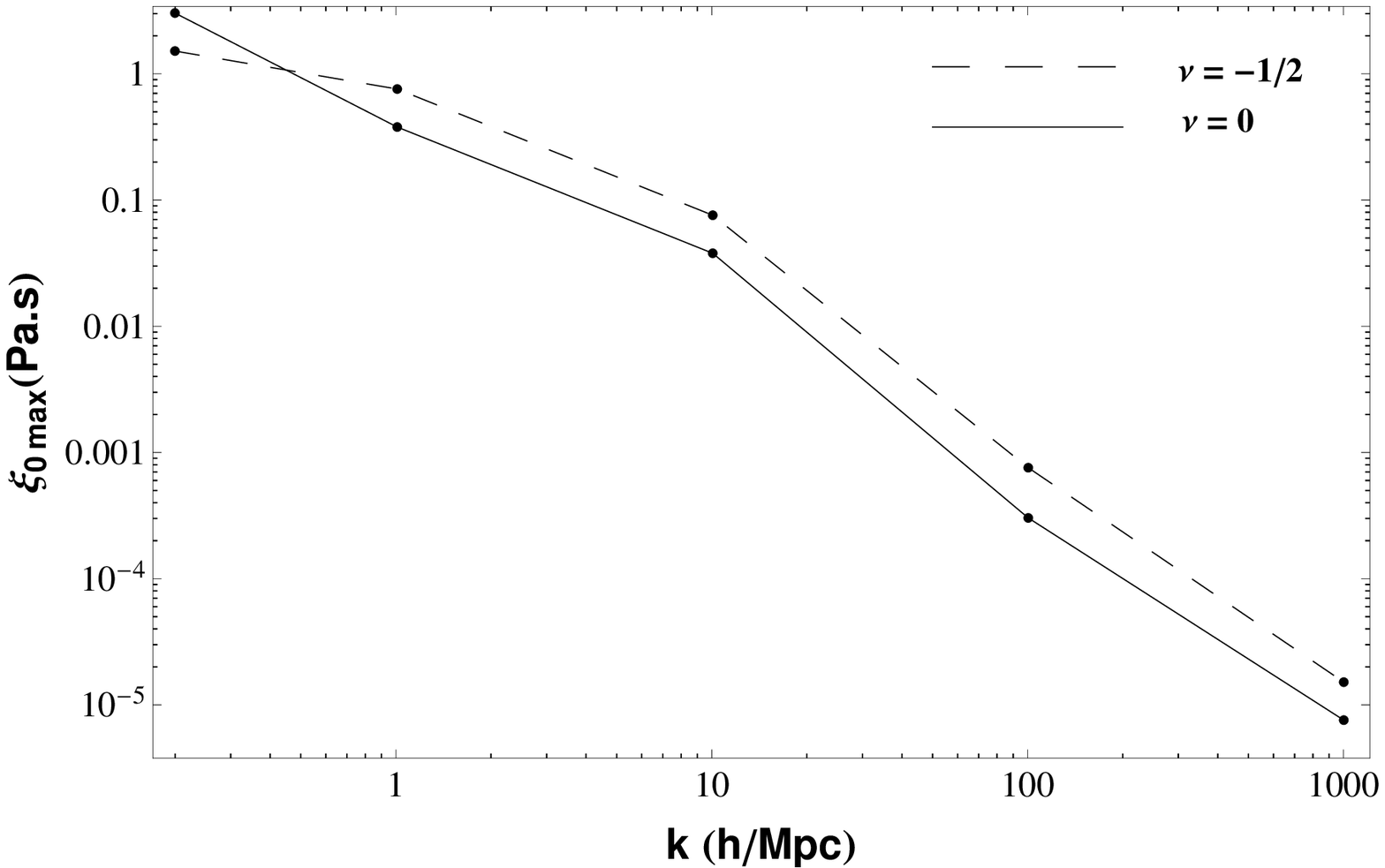}
\label{f3c}	
\caption{Maximum viscosity allowed following the requirement $\Delta_{\rm vCDM}(z)=\Delta_{\rm CDM}(z)$ for $z \geq z_{\rm nl}$. For $\tilde{\xi}>\tilde{\xi}_{max}$ and $\xi_0>\xi_{0max}$, structure formation is affected by viscous phenomena.}
\end{center}
\end{figure}

\section{Conclusions}

In order to investigate the dissipative properties of dark matter, we have proposed the $\Lambda$vCDM model, where cold dark matter is modeled as a dissipative fluid equipped with bulk viscous (negative) pressure (the vCDM fluid). Differently from UDM viscous models, where the viscous dark fluid is responsible for the accelerated expansion of the Universe, here we allow the existence of a nonvanishing cosmological constant. This means that the vCDM equation of state can be only slightly negative. This possibility is motivated by recent studies of galaxy clusters \cite{mariano}.

The free parameters of our model are $\Omega_{\Lambda}$ and the 
dimensionless viscosity parameter $\tilde{\xi}$. The limiting cases 
$\Omega_{\Lambda} = 0$ and $\tilde{\xi} = 0$ correspond to the viscous UDM model (the VDF model) and the $\Lambda$CDM model, respectively. The comparison with observations (see Fig.~1) has revealed that the unified model is strongly disfavored. The contour (at $2\sigma$) for the WiggleZ BAO data together with the position of the CMB first peak is decisive to put an upper limit on the dark matter viscosity $\tilde{\xi}$. For both vCDM dynamics studied here, namely models A ($\nu=0$) and B ($\nu=-1/2$), we obtain 
$\tilde{\xi}<0.2$ (at $2\sigma$) corresponding to $\xi_{0}<10^{7}$ Pa$\cdot$s in SI units. 

For the sake of comparison, the bulk viscosity coefficient of water at atmospheric pressure and at room temperature at $25\,^{\circ}{\rm C}$ is $2.5 \times 10^{-3}$ Pa$\cdot$s \cite{viscosityvalues}; thus we see that constraints from the dynamics of the homogeneous and isotropic background are rather poor. 

For the allowed (at $2\sigma$) model parameters, the $\Lambda$vCDM does not produce a large amplification of the integrated Sachs-Wolfe signal as in the viscous UDM model case. In the latter approach the background preferred data produces $Q>120\%$ \cite{dominik}. The current estimations of the error bars in the cross correlation of galaxy density and CMB temperature (used to measure the ISW effect) are still compatible with models where 
$Q \sim 100\%$ \cite{iswmore}. This means that amplifications of the order 
$Q \lesssim 40 \%$, as found here for the $\Lambda$vCDM model, are compatible with current observations. Radio surveys in the near future will reduce the error by a factor of $5$ and thus should be able to improve the limit to the 
$Q \sim 20 \%$ level (see figure 9 of \cite{ISWfuture}).

The $\Lambda$vCDM is much more tightly constrained by the analysis of the 
growth of subhorizon density perturbations. The production of small halos 
is a fundamental aspect of the hierarchical structure formation process. 
When even a small bulk viscous (negative) pressure is present, the growth of the smallest dark matter halos is suppressed as shown in Fig.~2. The suppression in model A (a constant bulk viscosity) is less severe than model B (which during the matter dominated phase is equivalent to a constant negative pressure). As seen in Fig.~2, vCDM is able to form galactic dark halos, within the hierarchical scenario, only if $\tilde{\xi} \ll 0.2$. For dwarf galaxy scales ($\sim 1$ kpc) the growth of the density contrast is similar to the standard CDM if the viscosity values are reduced to 
$\tilde{\xi} < 10^{-11}$. However, note that avoiding the formation of dwarf galaxies could provide a solution to the problem of the missing satellites in the standard CDM scenario \cite{missingsat}. Assuming that we have to guarantee, at least, the existence of $10$\,kpc galaxies, the viscosity parameter is set to $\tilde{\xi} < 10^{-9}$ or $\xi_{0}<10^{-1}$ Pa$\cdot$s in SI units. This result is even smaller than the limit $\xi < 7.38 \times 10^3$ Pa$\cdot$s, obtained in \cite{carlevaro}, where the authors study the Jeans mechanism for the bulk viscous fluid at the recombination era.

Since the parameter $\tilde{\xi}$ is of the same order of magnitude as the equation of state parameter today $w_{\rm v0}$ [see (\ref{wv0})], the results mentioned in the introduction, i.e $w_{\rm dm}\sim -0.2$ for the Coma cluster, are challenged by our analysis.

Our conclusions are limited by the linear analysis here performed. This formalism breaks down when $\Delta\sim 1$ and numerical simulations would be required to predict the final clustering patterns. The study of the nonlinear collapse of a vCDM component can be decisive to clarify whether virialized viscous structures survive at the end of the matter dominated epoch. A further limitation of our study relies on the assumption that the coefficient $\xi$ obeys the ansatz (\ref{xi}). A microscopic model for the bulk viscosity, as one finds in the ``dark goo'' model \cite{darkgoo}, would yield to a more appropriate coefficient $\xi$.

To summarize, we have shown that the strongest constraints on the 
dissipation of dark matter come from the study of hierarchical structure 
formation at the smallest observable scales. As dwarf galaxies are observed, 
their existence implies that $\xi_{0}<10^{-3}$ Pa$\cdot$s, a value as large 
as the bulk viscosity of water at normal conditions on Earth.

{\bf Acknowledgments}
 We thank J\'ulio Fabris, Friedrich Hehl and Winfried Zimdahl for comments and discussions.
HV thanks CNPq-Brazil for financial support. The authors acknowledge support from the DFG within the Research Training Group 1620 ``Models of Gravity''.

\end{document}